\documentclass{moriond}

\def\Journal#1#2#3#4{{#1} {\bf #2}, #3 (#4)}

\def\PRD{{\em Phys. Rev.} D}

\def\be{\begin{equation}}
\def\ee{\end{equation}}
\def\bea{\begin{eqnarray}}
\def\eea{\end{eqnarray}}

\usepackage{xcolor}

\usepackage{siunitx}

\newcommand{\Photo}{}

\begin{document}
\vspace*{4cm}
\title{STATUS OF QUBIC, THE Q\&U BOLOMETRIC INTERFEROMETER FOR COSMOLOGY}

\author{L. MOUSSET on behalf of the QUBIC collaboration}

\address{IRAP, Universit\'e de Toulouse, CNRS, CNES, UPS, Toulouse, France}

\maketitle{}\abstracts{The Q\&U Bolometric Interferometer for Cosmology (QUBIC) is a novel kind of polarimeter optimized for the measurement of the $B$-mode polarization of the Cosmic Microwave Background (CMB), which is one of the major challenges of observational cosmology. The signal is expected to be of the order of a few tens of nK, prone to instrumental systematic effects and polluted by various astrophysical foregrounds which can only be controlled through multichroic observations. QUBIC is designed to address these observational issues with a novel approach that combines the advantages of interferometry in terms of control of instrumental systematics with those of bolometric detectors in terms of wide-band, background-limited sensitivity.}

\section{Introduction}
The quest for $B$-mode polarization of the Cosmic Microwave Background (CMB) is among the major challenges of observational cosmology. Cosmic inflation predicts primordial scalar perturbations of the metric (density fluctuations), but also tensor perturbations, equivalent to primordial gravitational waves. These tensor modes should be imprinted in the CMB polarization fluctuations with a very specific signature: odd-parity patterns (curl term), known as $B$-modes~\cite{zaldarriaga}. The amplitude of the tensor modes, relative to the scalar modes, is parametrized by the so called tensor-to-scalar ratio $r$. The expected signal is weak, requiring high sensitivity detectors. In addition, astrophysical foregrounds produce non-primordial $B$-mode polarization, such as thermal emission from dust grains in the Galaxy.

The Q\&U Bolometric Interferometer for Cosmology (QUBIC) was designed to address the $B$-mode detection challenge~\cite{qubic1}. Characterization and calibration of a Technical Demonstrator (TD) started in 2018, at Astroparticle Physics \& Cosmology (APC) laboratory. In May 2021, the instrument has been sent to Argentina. A second calibration phase is undergoing and the instrument will be installed on the observation site in the next months. In Hamilton \textit{et al.}~\cite{qubic1}, we give forecasts for typical observations and measurements: with three years of integration on the sky and assuming perfect foreground removal as well as stable atmospheric conditions from our site in Argentina, our simulations show that we  can achieve a statistical sensitivity to the effective tensor-to-scalar ratio (including primordial and foreground $B$-modes) $\sigma(r)=0.015$.

\section{Control of systematic effects with QUBIC}
\begin{figure}[ht!]
	\centering
	\includegraphics[width=.48\linewidth]{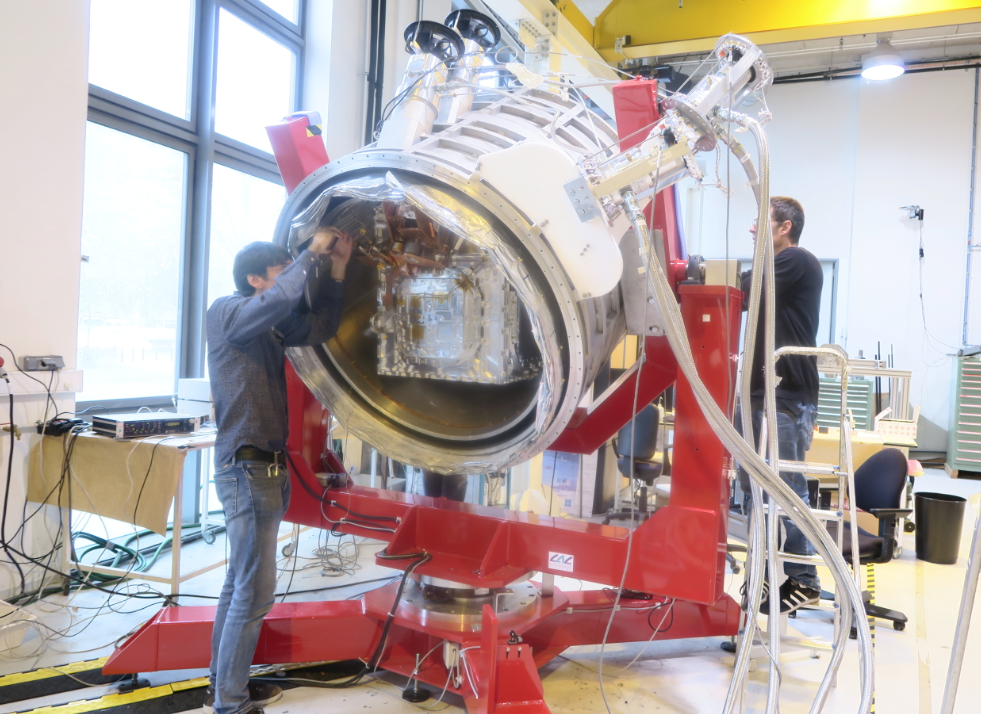}
	\includegraphics[width=.48\linewidth ]{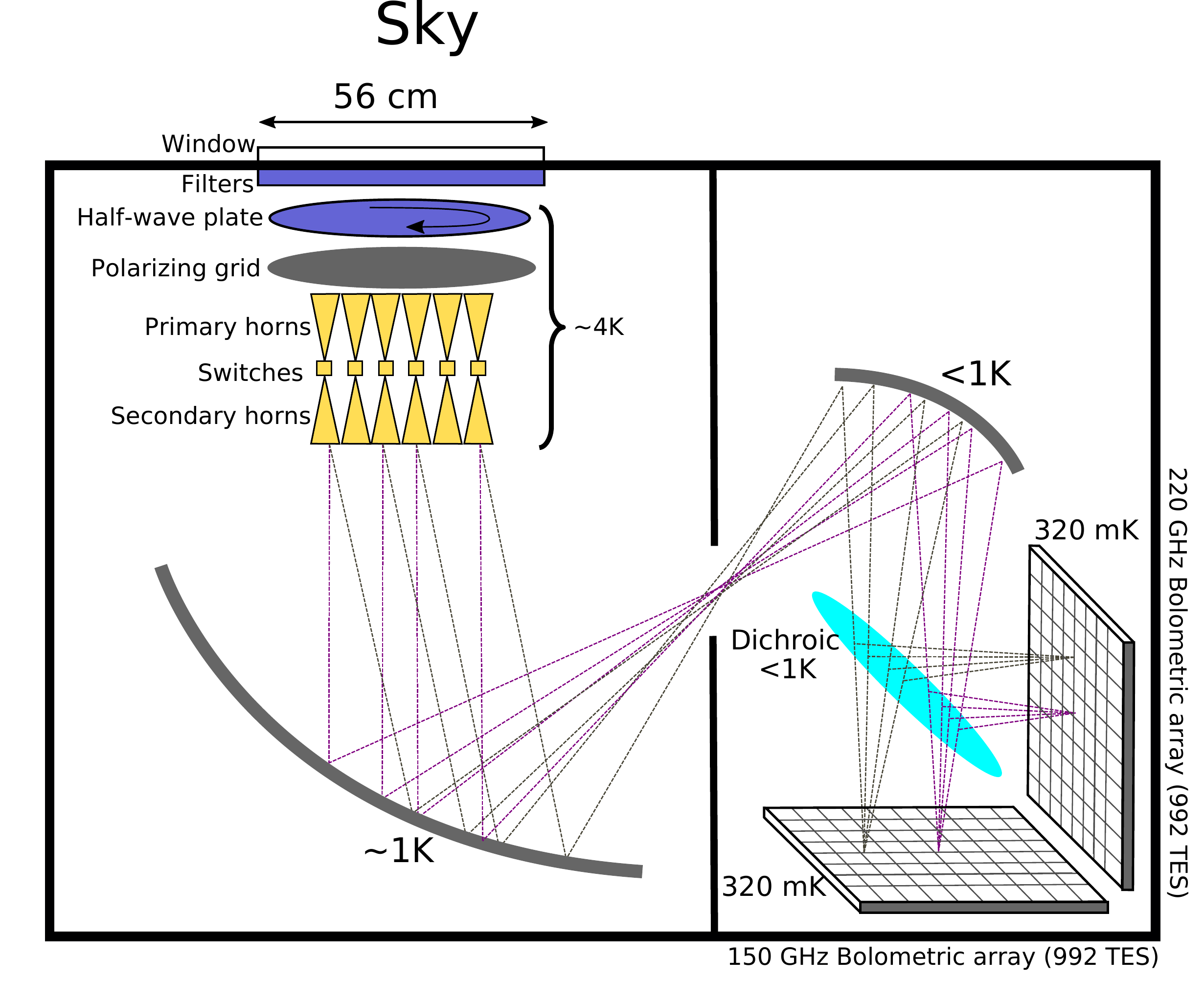}
	\caption{\textit{Left:} Picture of the QUBIC instrument at APC during the integration phase. \textit{Right:} Optical sketch of the instrument.}
	\label{Fig:qubic}
\end{figure}

As a bolometric interferometer, QUBIC combines the advantages of interferometry in terms of control of instrumental systematic effects with those of bolometric detectors in terms of wide-band, background-limited sensitivity. A picture of the instrument with a sketch of the optical design is shown in Figure~\ref{Fig:qubic}.

\subsection{Low cross-polarization}
QUBIC bolometers are full power detectors. The measurement of the polarization is only done by modulating the signal amplitude with a rotating Half-Wave Plate (HWP) and a fixed polarizing grid. Every optical element has its own systematic effects which could induce cross-polarization. This means that the two polarization directions $(x, y)$ of the signal can be mixed when interacting with the instrument and this is a major issue. The $x$-axis is defined as the transmission axis of the polarizing grid. By putting the HWP and the polarizing grid right after the window, the $x$ polarization is selected as early as possible. In this way, any cross-polarization occurring after the polarizing grid, for example generated by the horn-array, the mirrors or any reflection in the instrument, has no impact.

The modulation of the polarization was tested during the calibration phase and a very low cross-polarization was indeed detected. Those measurements are presented in Torchinsky \textit{et al.}~\cite{qubic3} and D'Alessandro \textit{et al.}~\cite{qubic6} and the cross-polarization contamination at 150~GHz is compatible with zero to within 0.6\%.

\subsection{Self-calibration technique}
\begin{figure}[ht!]
	\centering
	\includegraphics[width=.36\linewidth]{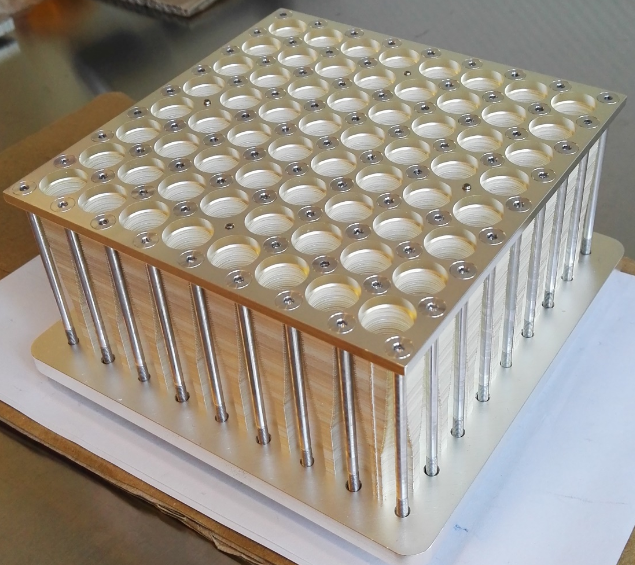}
	\includegraphics[width=.42\linewidth]{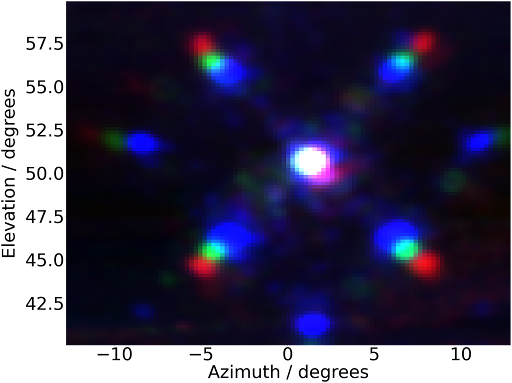}
	\caption{\textit{Left:} Picture of the TD horn-array ($8\times 8$ back-to-back horns), taken from Cavaliere \textit{et al.}~\protect\cite{qubic7}. \textit{Right:} Superimposition of three monochromatic beams measured in the laboratory at 130~GHz (red), 150~GHz (green) and 170~GHz (blue), taken from Torchinsky \textit{et al.}~\protect\cite{qubic3}.}
	\label{Fig:beam}
\end{figure}

Interferometry offers the possibility to self-calibrate the instrument systematic effects. This technique has been used for a long time in radio astronomy~\cite{cornwell}. Self-calibration relies on the concept of equivalent baselines, one baseline $b$ being formed with two horns. The full horn-array is shown in Figure~\ref{Fig:beam} (left).

The self-calibration technique is based on the fact that, in case of an ideal instrument without any systematic effect, equivalent baselines produce the same interference pattern on the focal plane, for an observation at infinity, in the Fraunhofer regime. Thus, by measuring the differences, one can calibrate the systematics of the instrument. The demonstration of this technique for QUBIC was done in Bigot-Sazy \textit{et al.}~\cite{bigot}. 

\section{Spectral imaging capability}

The instrument beam pattern, shown in Figure~\ref{Fig:beam} (right), is given by the geometric distribution of the horn-array. It contains multiple peaks whose angular separation is linearly dependent on the wavelength. As a result, and after a non-trivial map-making process, a bolometric interferometer such as QUBIC can simultaneously produce sky maps at multiple frequency sub-bands with data acquired over a single wide frequency band. The demonstration of this technique, called spectral imaging, and the characterization of its performance are presented in Mousset \textit{et al.}~\cite{qubic2}.

\subsection{Tested on end-to-end simulations}
We demonstrate spectral imaging capabilities by trying to recover the frequency dependence of the thermal galactic dust emission with simulated observations. We simulate an observation in a sky patch of 15~degree radius. The parameters of the pipeline are set in such a way that the simulated instrument has a single focal plane operating either at 150~GHz or at 220~GHz with a 25\% bandwidth each. From these wide-band TOD, we are able to reconstruct several numbers of sub-bands using spectral imaging. 

\begin{figure}[ht!]
	\centering
	\includegraphics[width=.9\linewidth ]{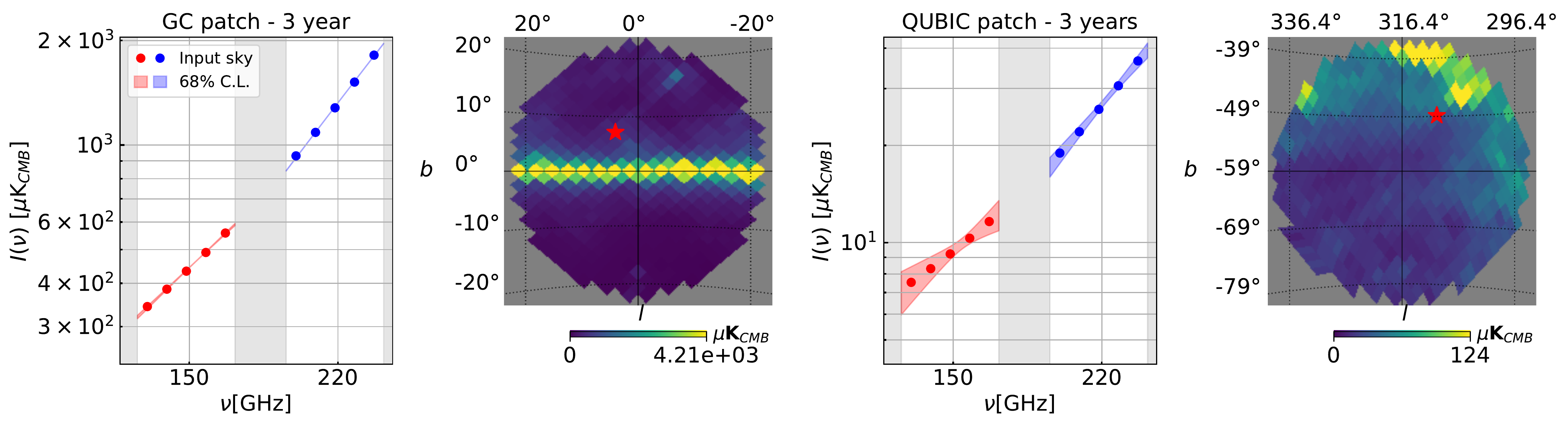}
	\caption{Intensity as a function of the frequency for $N_{\rm rec}=5$~sub-bands in each wide band at 150~(red) and 220~(blue)~GHz for a given pixel. The grey regions correspond to the unobserved frequencies outside our physical bands. Two sky pixels are shown as red stars, one in a patch centered at the Galactic center and one in the patch that QUBIC plans to observe centered in [0, -57~deg]. Red and blue dots: Input sky convolved with the instrument beam. In both cases are shown in light color the 68\%~CL regions for a modified black-body spectrum reconstructed with a MCMC from our simulated measurements and sub-band covariance matrices. Maps are in $\mu$K CMB and $N_{\rm side}=32$.}
	\label{Fig:pix-pix_plot}
\end{figure}

The reconstructed intensity as a function of frequency is studied in a given pixel. Figure~\ref{Fig:pix-pix_plot} shows the intensity of the input sky convolved with the instrument beam, and the reconstructed intensity for a given pixel, considering 5~sub-bands in each wide band at 150~(red) and 220~(blue)~GHz.

\subsection{First trial on real data}
\begin{figure}[ht!]
	\centering
	\includegraphics[width=.72\linewidth]{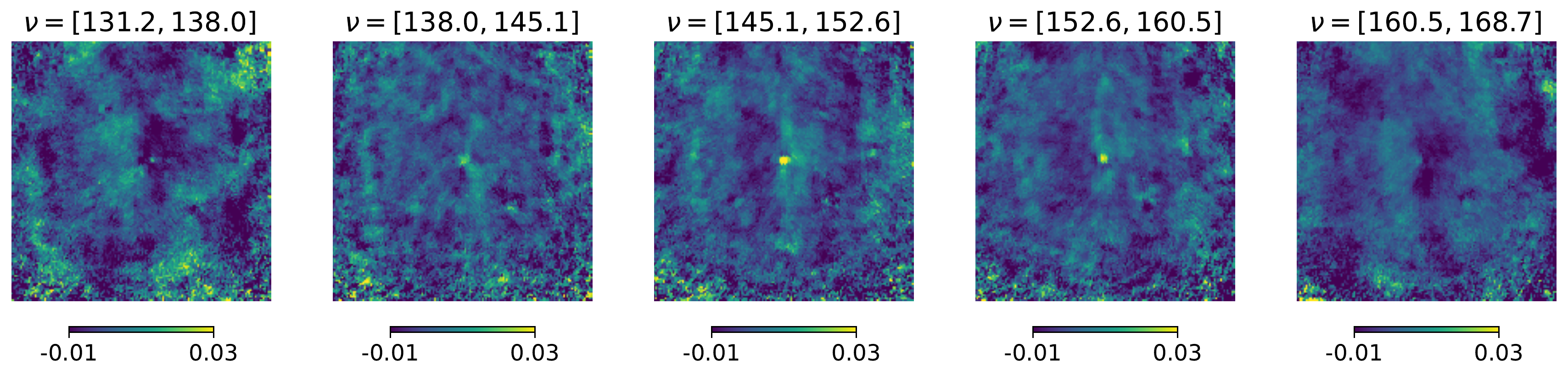}
	\includegraphics[width=.27\linewidth ]{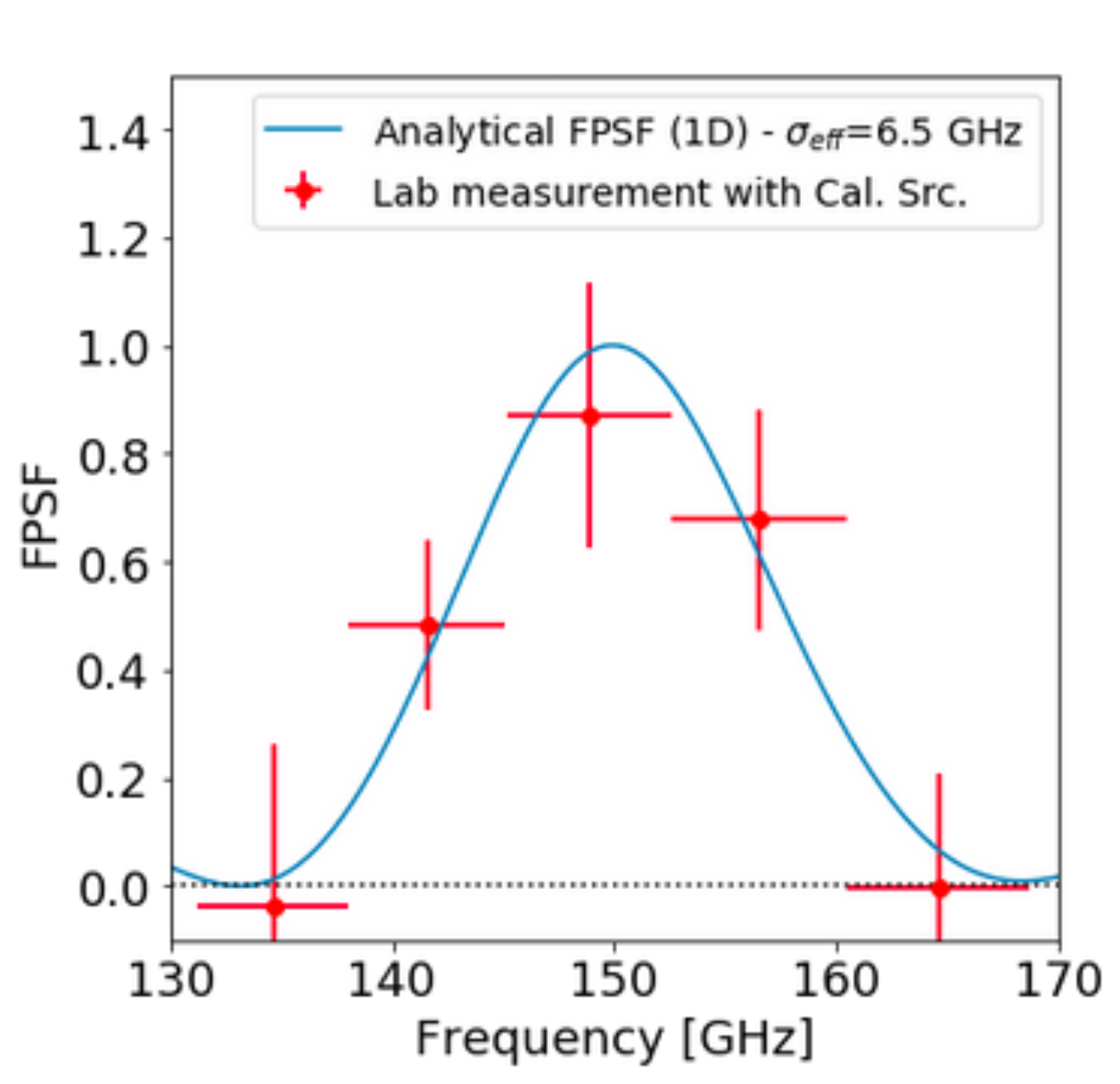}
	\caption{\textit{Left:} Calibration data with the source at 150 GHz projected on the sky using our map-making software to deconvolve from the multiple peaked synthesized beam and split the physical band of the instrument into 5 sub-bands. \textit{Right:} Measurement of the flux of the source in reconstructed sub-bands. The measurement (simple aperture photometry) in red is compared to the expected value spectrum in blue.}
	\label{Fig:real_data}
\end{figure}

Spectral imaging has been applied on real data for the first time during the calibration campaign at the APC laboratory. The QUBIC instrument was placed on an alt-azimuth mount in order to scan a calibration source tuned at 150~GHz (with 144~Hz bandwidth) and placed in the far field. The corresponding analysis is presented in Torchinsky \textit{et al.}~\cite{qubic3}. We performed a scan in azimuth and elevation with the instrument, obtaining a TOD for each bolometer. We then applied our spectral imaging map-making algorithm with five sub-bands to a selection of 26 bolometers that do not exhibit saturation. The synthesized beam for each bolometer is realistically modeled in our map-making through a series of Gaussian whose amplitude, width and locations are fit from a measured map of the synthesized beam for each bolometer (see Figure 20 from Torchinsky \textit{et al.}~\cite{qubic3} for an example). We were able to reconstruct a map of the point-like artificial calibration source as well as its location in frequency space. In Figure~\ref{Fig:real_data}, we show the reconstruction onto 5~sub-bands. The expected point-source shape is clearly visible in the central frequency sub-band containing the emission frequency of the source at 150~GHz, it is fainter in adjacent bands, and not visible in the furthest bands. On the right, we show the detected amplitude in the central pixel as a function of the frequency. The measurement in red is compared to the expected value spectrum in blue.

\section{Conclusion and perspectives}

The QUBIC Technical Demonstrator was successfully tested at APC laboratory and sent to Argentina to be installed on the observation site in the following year. The QUBIC instrument relies on an innovative design which allows a very low cross-polarization, a high control of systematic effects thanks to self-calibration and a spectroscopic capability. Spectral imaging technique was demonstrated on end-to-end simulations and applied to calibration data successfully. Testing it on real astrophysical sources is the next determining step. 


\end{document}